# Introduction to Mechanics and Structures

*M. Scapin*
Politecnico di Torino, Department of Mechanical and Aerospace Engineering, Corso Duca degli Abruzzi, 24, 10129, Turin, Italy

**Abstract**
This work provides a comprehensive overview of the fundamental concepts in continuum mechanics, focusing on the behaviour of materials under mechanical loads. It discusses the distinction between elastic and plastic, highlighting their atomic origins and macroscopic implications. Elastic behaviour is examined via Hooke's law and constitutive matrices, while plasticity is treated through yield surfaces, flow rules, and hardening laws, including isotropic and kinematic hardening. In addition, the theoretical foundations and design principles of pressure vessels and thin axisymmetric shells, focusing on their mechanical behaviour under internal or external pressure, is discussed. The analysis is based on shell theory, assuming thin walls and axisymmetric geometry, which simplifies the stress distribution into membrane stresses. The work also addresses buckling phenomena under external pressure, secondary stresses at geometric discontinuities, and design provisions from the EN 13445 standard.

## 1 Basics of continuum mechanics

When a material is subjected to external forces, the atoms inside it are displaced from their original equilibrium positions. This displacement manifests itself in a deformation of the body. The type of deformation determines the mechanical behaviour of the material. Two broad categories of deformation exist: reversible and irreversible.

– Reversible deformations disappear when the load is removed. These are defined as elastic deformations and are governed by the atomic bonding and the stiffness of the material.

– Irreversible deformations remain after the load is removed. These are defined as plastic deformations, in which atomic rearrangements prevent the material from returning to its initial state.

Another important distinction concerns time dependence. Deformations can be time-independent (purely elastic or plastic) or time-dependent. The latter occur in materials with viscous behaviour, such as polymers at elevated temperatures. These are generally termed viscoelastic or visco-plastic responses.

Understanding the differences between these behaviours is fundamental for predicting how structural components respond under mechanical loading.

### 1.1 State of stress

Whenever a body is loaded externally, internal forces are distributed within its volume to maintain equilibrium. If we consider an imaginary cut inside the body, the particles on either side of the cut exert forces on each other across the surface. The measure of this internal force per unit area is called the stress vector. It depends not only on the location within the body but also on the orientation of the surface considered. On an element of area $\Delta A$ containing a point $P$, with normal vector $n$, the force distribution is equipollent to a contact force $\Delta F$ exerted at point P and surface moment $\Delta M$ (see Fig. 1 left). When $\Delta A$ becomes small and tends to zero the ratio $\Delta F/\Delta A$ becomes $dF/dA$ and the couple stress vector $\Delta M$ vanishes. The resultant vector $dF/dA$ is defined as stress vector at the point $P$ associated with a plane with a normal vector $n$: This means that the stress vector depends on its location in the body (i.e. which is the point P) and the orientation of the plane on which it is acting. The stress state at a point in the body is then completely defined by all the stress vectors associated with all the planes (infinite in number) that pass through that point.

However, according to Cauchy's stress theorem, merely by knowing the stress vectors on three mutually perpendicular planes, the stress vector on any other plane passing through that point can be found through coordinate transformation equations. This theorem provides the foundation for moving from a vectorial to a tensorial representation of stress.

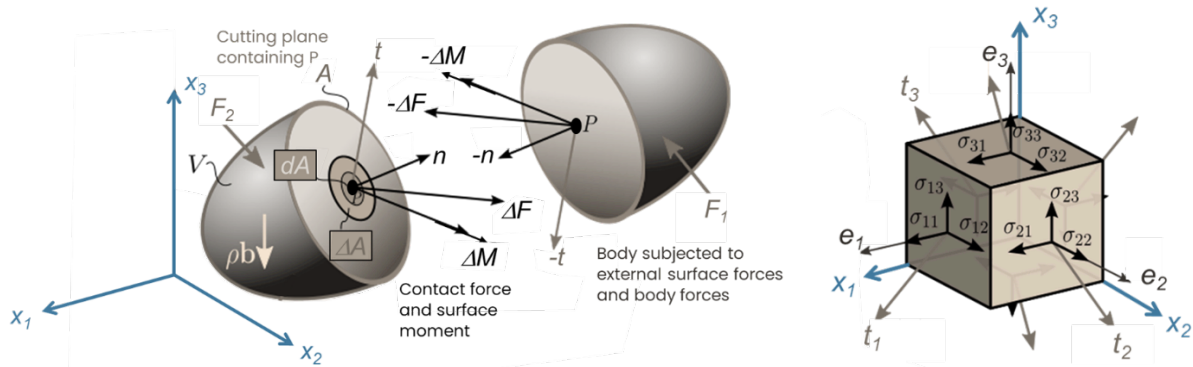

**Fig. 1:** Stress vectors (left) and Cauchy stress components (right) in a continuum body.

This means that the stress state at a point can be fully described using the stress tensor, according to the Eq. (1), in which each element of the tensor represents a normal or shear stress component acting on one of the three coordinate planes (see Fig. 1 right). Due to equilibrium and compatibility conditions, the stress tensor is symmetric, leaving only six independent components. The diagonal components are the normal stresses acting perpendicular to each coordinate plane, while the off-diagonal terms represent shear stresses acting tangentially.

$$t = \begin{bmatrix} \sigma_{11} & \sigma_{21} & \sigma_{31} \\ \sigma_{12} & \sigma_{22} & \sigma_{32} \\ \sigma_{13} & \sigma_{23} & \sigma_{33} \end{bmatrix} n \qquad (1)$$

A change of coordinate system modifies the tensor components, but it still describes the same stress state. The relation between the representation in the two different coordinate systems can be stated using the transformation matrix.

Depending on the orientation of the plane under consideration, the stress vector may not necessarily be perpendicular to that plane, i.e. parallel to $n$, and can be resolved into two components (see Fig. 2 left):

- one normal to the plane, called normal stress,
- and the other parallel to this plane, called shear stress.

Mohr's circle is a graphical tool (see Fig. 2 right) that provides valuable insight into how stresses transform when the coordinate system is rotated. The abscissa and ordinate ($\sigma_n, \tau_n$) of each point on a circle are the magnitudes of the normal stress and shear stress components, respectively. Each circle is the locus of points that represent the stress state on an individual plane (perpendicular to the third direction) at all its orientation.

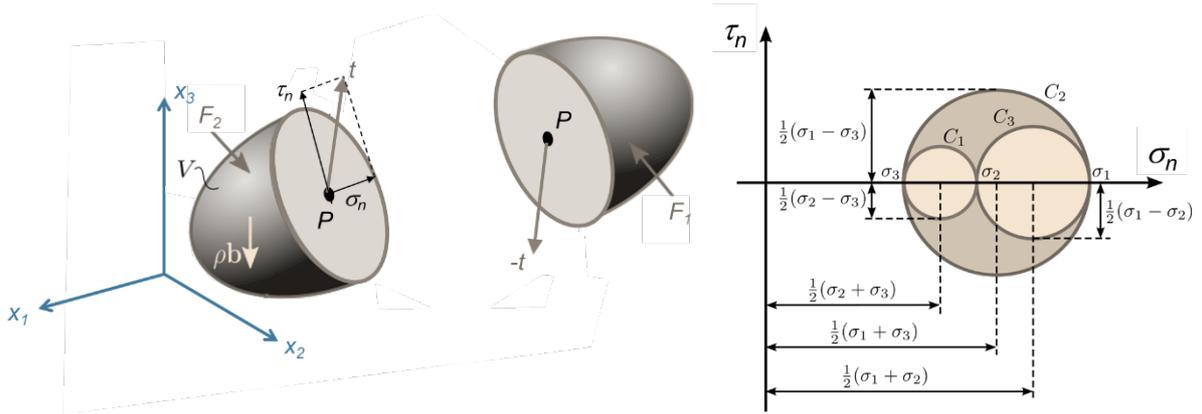

**Fig. 2:** Normal and shear stresses (left) and Mohr's circles representation of the stress state (right).

Principal stresses are special stress values corresponding to orientations where shear stresses vanish (see Fig. 3). In this coordinate system, the stress tensor reduces to a diagonal form expressed by Eq. (2) and the principal stresses σ₁,σ₂,σ₃ are obtained by solving the eigenvalue problem reported in Eq. (3). For each eigenvalue, there is a non-trivial solution which is a principal direction or eigenvector defining the plane where each principal stress acts. The principal stresses and principal directions characterize the stress at a point and are independent of the orientation.

$$\boldsymbol{t} = \begin{bmatrix} \sigma_1 & 0 & 0 \\ 0 & \sigma_2 & 0 \\ 0 & 0 & \sigma_3 \end{bmatrix} \boldsymbol{n} \qquad (2)$$

$$(\boldsymbol{\sigma} - \sigma_p \boldsymbol{I})\boldsymbol{n} = \boldsymbol{0} \qquad (3)$$

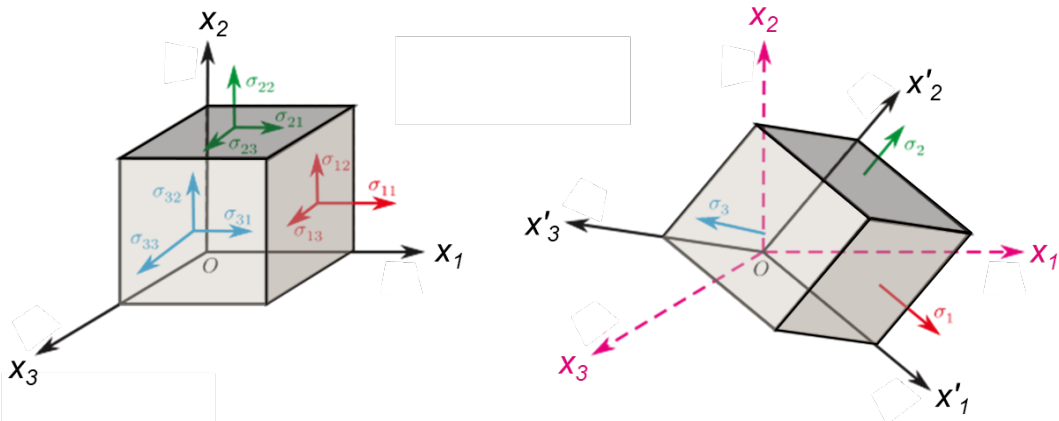

**Fig. 3:** Cartesian (left) and principal (right) stresses and directions.

## 1.2 State of strain

If a component is stressed, points within it are displaced. To describe the deformation changes of distances and angles between points have to be looked at.

An arbitrary deformation with small strains of a material element can be described– analogous to the stress– by a tensor, the strain tensor (Eq. (4)). To calculate the strain tensor, we chose a coordinate system that is fixed in space and consider the displacement of material points in this system (Lagrangian point of view). As for stresses, the strain tensor can be diagonalised, yielding its principal components, which represent the normal strains acting along the principal directions where shear contributions vanish.

$$\boldsymbol{\varepsilon} = \begin{bmatrix} \varepsilon_{11} & \varepsilon_{21} & \varepsilon_{31} \\ \varepsilon_{12} & \varepsilon_{22} & \varepsilon_{32} \\ \varepsilon_{13} & \varepsilon_{23} & \varepsilon_{33} \end{bmatrix} \quad (4)$$

### 1.3  Deviatoric and hydrostatic tensor

Both the stress and the strain tensors can be decomposed into two fundamental contributions:

- The first is the hydrostatic part, corresponding to the mean value of the normal components of the tensor. It is associated with a change in volume of the body. In terms of stresses, the hydrostatic component coincides with minus the pressure, while for strains it corresponds to the volumetric (or bulk) strain, defined as the relative variation of volume and expressed by the first invariant of the strain tensor, i.e. its trace.

- The second contribution is the deviatoric part, obtained by subtracting the hydrostatic component from the total tensor. This part is traceless and is solely responsible for distortional effects: it modifies the shape of the body without altering its volume. Consequently, the deviatoric stress produces distortional strain, and the corresponding deviatoric strain tensor characterises the distortion of the body at constant volume.

## 2  Elasticity

Elasticity in the field of mechanics originates from the fundamental interactions between atoms within a material. When an external load is applied, atomic bonds are slightly displaced from their equilibrium positions, and the resulting interatomic forces act to restore the original configuration once the load is removed. This reversible relationship between stress and strain constitutes the essence of elastic behaviour (Fig. 4 left).

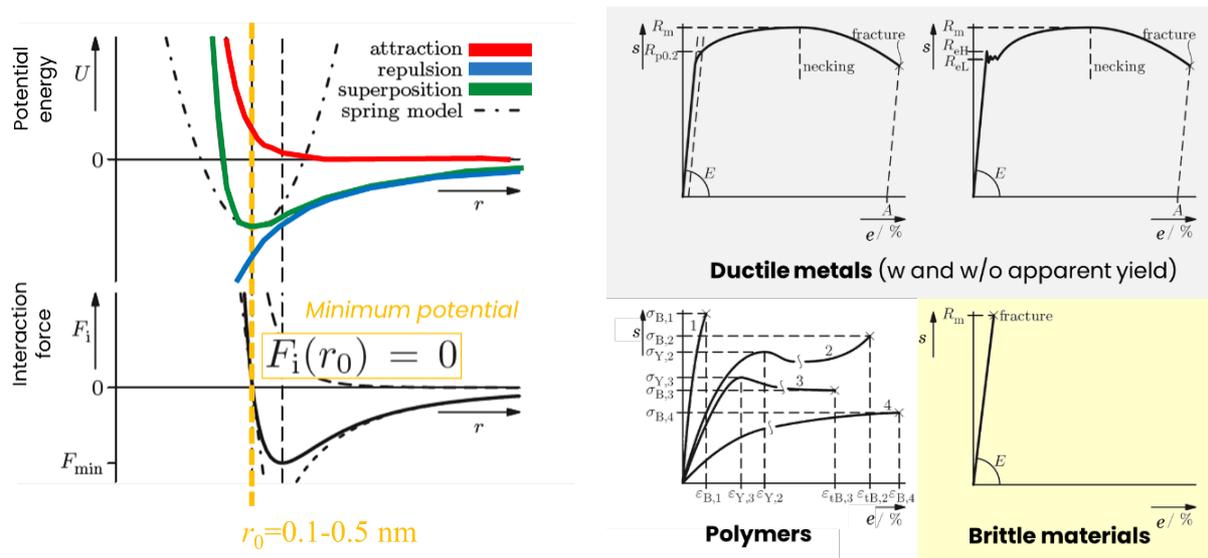

**Fig. 4:** Elasticity: Atomic interaction (left); Engineering stress-strain curve for several class of materials (right).

At the macroscopic level, such phenomena are mathematically represented through the constitutive matrix, which linearly relates the components of stress and strain according to Hooke's law. In the most general anisotropic case, the constitutive matrix is fully populated and characterized by twenty-one independent elastic constants. For orthotropic materials, where three mutually orthogonal planes of material symmetry exist, the number of independent constants reduces to nine. In the isotropic case, which describes materials whose mechanical properties are identical in all directions, the constitutive behaviour is further simplified and governed by only two independent parameters: the Young's modulus and the Poisson's ratio. When thermal effects are considered, an additional term accounting for thermal strains is introduced, leading to a thermoelastic formulation in which stress depends not only on mechanical strain but also on temperature variations through the coefficient of thermal expansion. In isotropic materials, normal stresses are associated exclusively with volumetric (dilatational) deformations, whereas shear stresses are related solely to distortional (deviatoric) deformations, reflecting the clear decoupling between volumetric and shear responses (Eq. (5)).

$$\{\varepsilon\} = [S]\{\sigma\} + \alpha \Delta T \begin{Bmatrix} 1 \\ 1 \\ 1 \\ 0 \\ 0 \\ 0 \end{Bmatrix}; [S] = \begin{bmatrix} \frac{1}{E} & -\frac{\nu}{E} & -\frac{\nu}{E} & 0 & 0 & 0 \\ -\frac{\nu}{E} & \frac{1}{E} & -\frac{\nu}{E} & 0 & 0 & 0 \\ -\frac{\nu}{E} & -\frac{\nu}{E} & \frac{1}{E} & 0 & 0 & 0 \\ 0 & 0 & 0 & \frac{2(1+\nu)}{E} & 0 & 0 \\ 0 & 0 & 0 & 0 & \frac{2(1+\nu)}{E} & 0 \\ 0 & 0 & 0 & 0 & 0 & \frac{2(1+\nu)}{E} \end{bmatrix} \quad (5)$$

## 3 Plasticity

Plasticity represents the branch of solid mechanics that deals with the permanent deformation of materials under load, i.e., when the relationship between stress and strain ceases to be linear and reversible. Unlike the elastic regime, where atoms return to their original equilibrium positions once the external load is removed, plastic deformation is characterized by an irreversible rearrangement of atomic positions. During plastic deformation, atoms within the crystalline lattice overcome the energy barriers associated with the potential wells of neighbouring atoms and move to new equilibrium configurations. As a consequence, the material does not recover its initial geometry after unloading. This atomic-scale rearrangement lies at the origin of the macroscopic phenomenon known as plastic flow. A clear example is provided by the condition of pure shear, in which both the stress and the strain tensors reduce to their deviatoric part. In this case, the body experiences shape changes only, with no volumetric variation.

### 3.1 Engineering stress-strain curve

The tensile test is the most fundamental and widely used experimental procedure to study the plastic behaviour of materials. In this test, a specimen—typically of standardized geometry—is subjected to a uniaxial tensile load, while both the applied force and the corresponding elongation are measured continuously. The results are then represented in terms of engineering stress and engineering strain.

The engineering stress $s$ is defined as the ratio between the applied force $F$ and the initial cross-sectional area $A_0$ of the specimen. The engineering strain $e$ is defined as the relative change in length with respect to the initial gauge length $L_0$ (Eq. 6).

$$s = \frac{F}{A}; \; e = \frac{\Delta L}{L_0} \quad (6)$$

These definitions are appropriate for small deformations, where changes in geometry are negligible. However, as deformation progresses into the plastic regime, the cross-sectional area and the gauge length vary significantly, leading to a progressive discrepancy between the engineering quantities and the actual state of stress and strain within the material. The typical engineering stress–strain curve obtained from a tensile test begins with a linear region, corresponding to elastic behaviour and governed by Hooke's law. Beyond a certain point—known as the yield point or yield strength—the curve deviates from linearity, marking the onset of plastic deformation. From this stage onward, the material exhibits a permanent deformation even after unloading. As loading continues, the curve typically reaches a maximum corresponding to the ultimate tensile strength (UTS). After this point, the observed engineering stress begins to decrease until fracture occurs. It is important to emphasize that this post-UTS softening does not necessarily indicate a reduction in the material's intrinsic strength, but rather a geometric effect associated with the phenomenon of necking. The shape of the stress–strain curve and the extent of the plastic region vary significantly depending on the material's nature (Fig. 4 right). Ductile materials, such as most metals, display a well-developed plastic region, allowing large permanent deformations before fracture. Their stress–strain curve shows a gradual transition from elastic to plastic behaviour, often with a distinct yield plateau and significant strain hardening. Brittle materials, such as ceramics and many glasses, exhibit almost no plastic deformation: they fail shortly after reaching the elastic limit, and their stress–strain curves terminate abruptly. Polymers, on the other hand, present a more complex, highly non-linear mechanical response even in the early stages of deformation, due to the viscoelastic and visco-plastic nature of their molecular chains.

### 3.2 True stress-strain curve

As deformation proceeds, especially beyond the elastic limit, the use of engineering quantities becomes increasingly inadequate. This is because engineering stress and strain are based on the initial geometry, which no longer represents the actual state of the specimen. To accurately describe the real mechanical response, it becomes necessary to introduce the true stress and true strain, defined on the current configuration of the material.

Under the assumption of volume constancy during plastic deformation which holds approximately for most metals, a simple relationship can be established between engineering and true quantities (Fig. 5 left). The true stress is defined as the instantaneous load divided by the instantaneous cross-sectional area. Similarly, the true strain is defined as the natural logarithm of the ratio between the current length and the initial length (Eq. (7)). This logarithmic measure arises naturally when considering strain as the cumulative sum of infinitesimal elongations, each referred to the current configuration. It thus provides an additive and physically consistent measure of deformation even for large strains.

$$\varepsilon = \frac{dl}{l} = \ln(1+e)\,; \sigma = \frac{F}{A} = s(1+e) \tag{7}$$

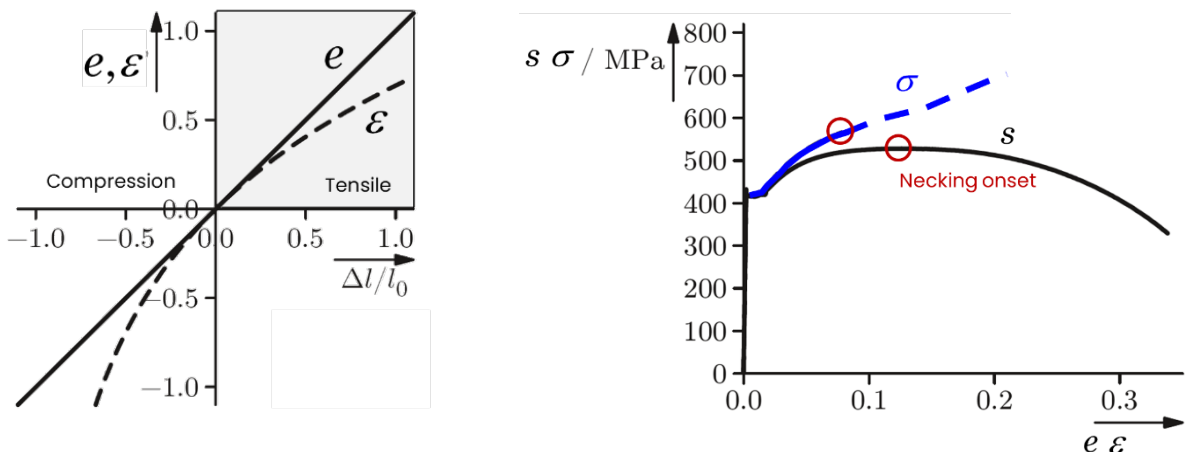

**Fig. 5:** Engineering strain vs true strain (left); Engineering vs. true stress-strain curves (right).

The true stress–strain curve thus provides a more faithful representation of the material's intrinsic response, showing a continuous hardening behaviour up to the onset of necking (Fig. 5 right).

## 3.3 Necking

As plastic deformation continues under uniaxial tension, the specimen eventually reaches a point where the load-carrying capacity begins to localize in a small region, forming a visible constriction known as the neck. Necking represents a transition from a homogeneous deformation state to a non-uniform one. The onset of necking can be understood by analysing the equilibrium between the rate of strain hardening and the rate of geometrical softening. Up to the necking point, strain hardening predominates: as the material is stretched, its ability to carry additional load increases due to dislocation interactions and the evolution of the microstructure. However, once the reduction in cross-sectional area becomes dominant, the applied load required to maintain equilibrium decreases, leading to instability and localization.

From a mechanical standpoint, necking corresponds to the point of maximum load on the engineering stress–strain curve. Before necking, the deformation is homogeneous along the gauge length: the stress and strain are uniformly distributed, and the uniaxial stress state assumption holds. Once necking begins, the stress distribution becomes non-uniform, and the local triaxiality of stress increases within the necked region due to geometric constraints and lateral contraction (Fig. 6 left).

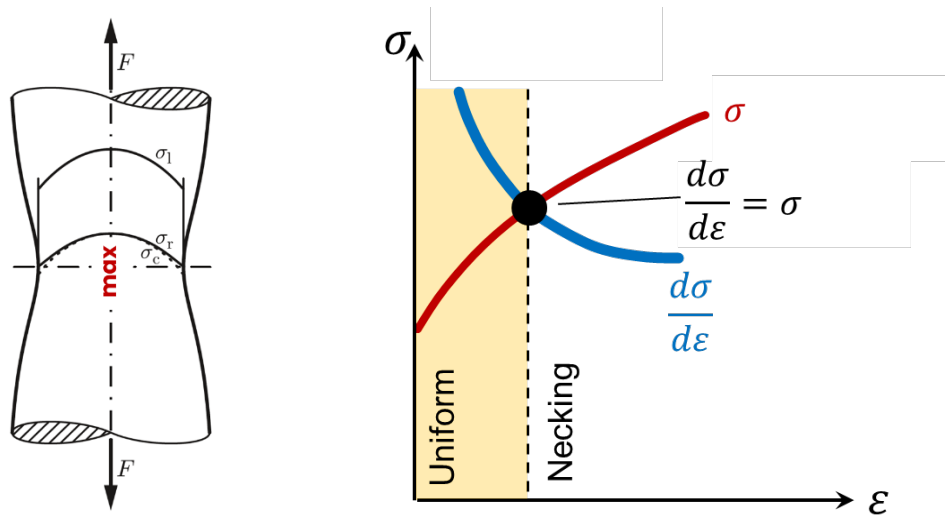

**Fig. 6:** Stress state during necking (left); Considère criterion (right).

The Considère criterion provides a simple condition for the onset of necking (Fig. 6 right). Considering the true stress–true strain curve, necking begins when the increase in load due to strain hardening exactly compensates the decrease due to the reduction in cross-sectional area. This equality states that necking starts when the rate of strain hardening equals the current true stress. Before this point, strain hardening dominates, ensuring uniform deformation; beyond it, any additional strain leads to a decrease in load capacity and the development of a localized neck. The physical interpretation of this criterion is that it marks the limit of stability of uniform plastic flow. Once the rate of geometrical softening exceeds the rate of hardening, the system can no longer sustain homogeneous deformation, and the material response becomes inherently localized.

## 3.4 Constitutive tools for the analysis of plastic behaviour

The transition from elastic to plastic behaviour, and the subsequent evolution of plastic deformation, can be effectively described through a set of tools that collectively form the theoretical framework of plasticity. These include the yield surface, which defines the onset of plastic flow; the flow rule, which determines the direction of plastic strain increments; and the hardening law, which governs the evolution of the material's resistance as deformation progresses. Together, these concepts provide a comprehensive representation of how materials respond under complex stress states, beyond the uniaxial conditions typical of tensile testing.

From a practical standpoint, these tools are indispensable for the analysis and numerical simulation of material behaviour under complex loading. In computational mechanics, they form the basis of elasto-plastic constitutive models, implemented within finite element formulations to predict stress–strain responses under arbitrary boundary conditions.

### 3.4.1 Yield surface

The yield surface (or yield criterion) defines the limit between elastic and plastic behaviour in stress space. Physically, it represents the locus of stress states at which the material begins to yield, that is, where irreversible plastic deformation initiates. Within the yield surface, all stress states correspond to purely elastic responses; once the stress state reaches the surface, plastic flow begins, and the material can no longer sustain additional stress increases without undergoing permanent deformation.

For ductile metals, the most widely used criterion is the von Mises yield criterion, which assumes that yielding depends solely on the deviatoric component of the stress tensor, i.e., the part responsible for shape change rather than volume change. This implies that yielding is independent of the hydrostatic stress, consistent with the observation that metals are essentially incompressible during plastic flow. From the geometrical point of view, in the three-dimensional principal stress space, the von Mises yield surface is represented by a cylinder aligned with the hydrostatic axis (Fig. 7 left), since the criterion is insensitive to the hydrostatic component of stress. When visualized in the deviatoric plane, the cross-section of this cylinder is a circle. If one considers the plane ($\sigma_3 = 0$), corresponding to a two-dimensional stress state, the projection of the yield surface becomes an ellipse.

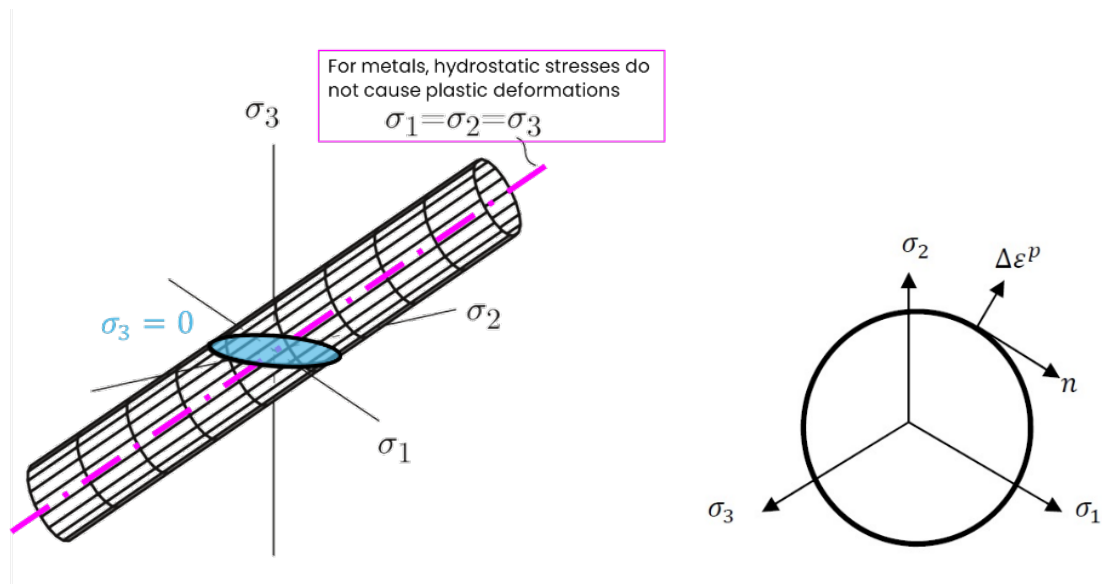

**Fig. 7:** Von Mises yield surface (left); Flow rule (right).

### 3.4.2 Flow rule

Once the yield condition is satisfied and plastic deformation occurs, it becomes necessary to define how the plastic strain increments develop. This is the role of the flow rule, which establishes the direction of plastic strain rates in stress space. In the case of associated flow, the condition of normality is valid: the plastic strain rate is directed along the normal to the yield surface in stress space. Since most yield criteria, such as the von Mises criterion, depend only on the deviatoric part of the stress tensor, the resulting plastic strain increments are also proportional to the deviatoric stress components (Fig. 7 right). This means that plastic flow is driven purely by distortional (shape-changing) effects, while the hydrostatic part of the stress — which represents a uniform pressure — does not contribute to plastic deformation.

### *3.4.3 Hardening behaviour*

While the yield surface defines the initial condition for yielding, real materials exhibit the ability to evolve their resistance to plastic deformation as plastic strain accumulates. This phenomenon, known as hardening, reflects the microscopic processes of dislocation interaction, multiplication, and rearrangement that occur during plastic flow. From a constitutive perspective, hardening is modelled by allowing the yield surface to evolve in stress space as a function of the internal variables introduced earlier. Two main types of hardening are distinguished: isotropic hardening and kinematic hardening. In isotropic hardening (Fig. 8 left), the yield surface expands uniformly in all directions as plastic deformation increases, preserving its shape and centre. This reflects an overall increase in the yield stress of the material due to the accumulation of dislocations which impede further motion. In kinematic hardening (Fig. 8 right), instead of expanding, the yield surface translates in stress space without changing its size or shape. This translation reflects the Bauschinger effect, i.e., the experimentally observed reduction in yield stress when the loading direction is reversed after plastic deformation. More advanced constitutive models combine both isotropic and kinematic.

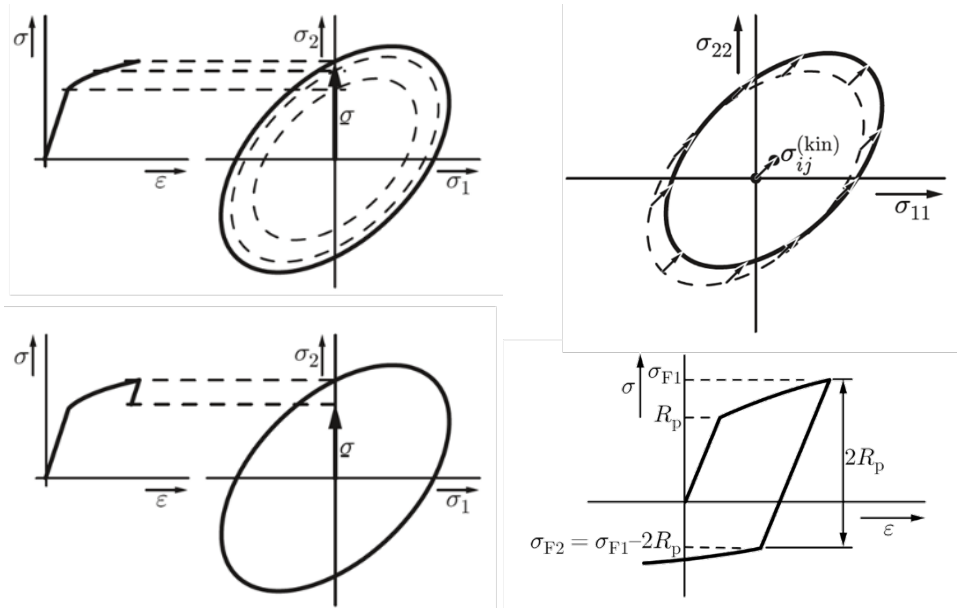

**Fig. 8:** Isotropic (left) and kinematic (right) hardening.

## 4    Pressure Vessels and Thin Axisymmetric Shells – Theoretical Foundations

Pressure vessels are structural components designed to contain fluids under pressure while maintaining their integrity and preventing leakage. They are fundamental in chemical, nuclear, and mechanical engineering applications, where they serve as reactors, heat exchangers, storage tanks, or containment systems. From a mechanical standpoint, a pressure vessel is a thin-walled, closed container that can sustain internal or external pressure, with stresses distributed primarily along its surface.

Although vessels may take various forms—cylindrical, spherical, conical, or combinations thereof—the most common configuration is a cylindrical shell with curved end caps (heads). The geometry of such components is typically axisymmetric, meaning that all geometrical, material, and loading properties are invariant with respect to rotation around a central axis. This symmetry significantly simplifies the stress analysis, allowing the use of shell theory under axisymmetric conditions.

Pressure vessels are characterized by a thin-wall assumption, where the wall thickness $h$ is small compared to the other dimensions. This simplification enables the analysis to be based on the membrane theory of shells, which neglects bending effects and assumes that stresses are uniformly distributed through the thickness. Under these conditions, the vessel mainly experiences in-plane stresses, known as membrane stresses, which act tangentially to the surface.

When pressure acts on a flat plate, out-of-plane bending is necessary to equilibrate the load. In contrast, curved shells can sustain transverse loads primarily through membrane forces, thanks to their initial curvature. Consequently, thin shells are highly efficient structural forms capable of withstanding significant pressure with minimal material use, a principle that underlies their widespread adoption in engineering design.

## 4.1 Geometry of Axisymmetric Shells

The geometry of an axisymmetric shell can be described by its middle surface, an imaginary surface equidistant from the inner and outer surfaces of the wall. The thickness of the shell is measured normally to this surface. For axisymmetric shapes, the middle surface can be generated by rotating a plane curve—called the meridian or generating line—around the axis of symmetry. The trajectories of points on the generating line form parallels, which are circular sections perpendicular to the axis (Fig. 9).

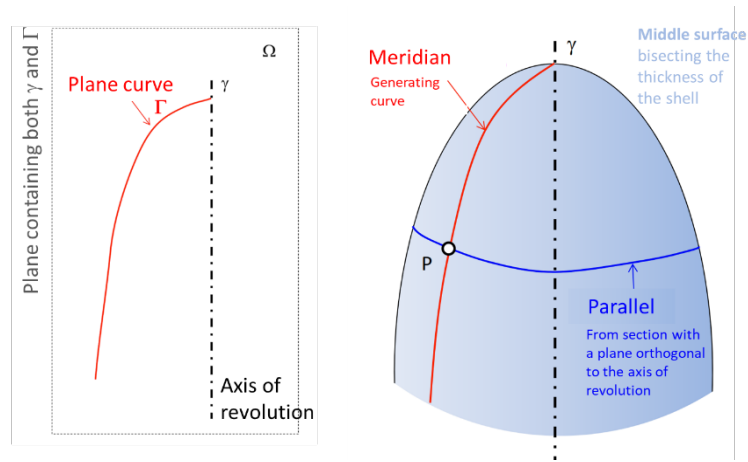

**Fig. 9:** Shell of revolution.

At any point $P$ on the middle surface, two principal curvatures define the geometry (Fig. 10). The curvature in the meridional plane (plane containing the axis) is associated with the meridional radius of curvature $r_\varphi$, while the curvature in the circumferential direction corresponds to the circumferential radius of curvature $r_\theta$. These two principal curvatures are orthogonal and uniquely define the local geometry of the surface. For a cylindrical shell, $r_\varphi = \infty$ and $r_\theta = R$; for a spherical shell, $r_\varphi = r_\theta = R$.

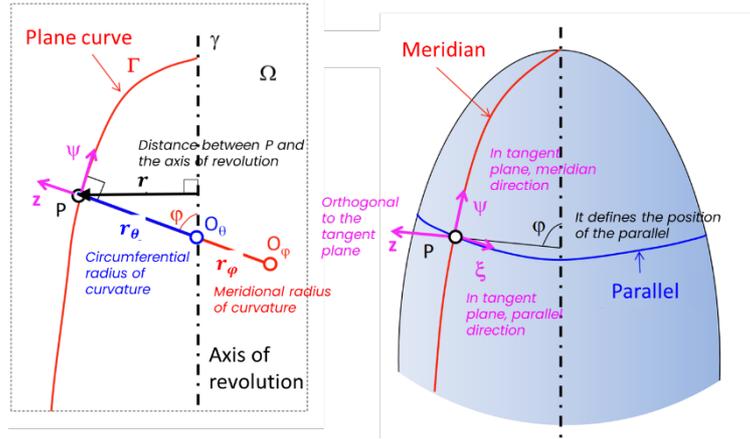

**Fig. 10:** Curvatures and curvilinear coordinate system cantered in the point P.

This geometrical framework provides the basis for establishing the local coordinate system used in shell analysis, typically defined by the meridional (φ), circumferential (θ), and normal (n) directions.

## 5  Membrane State of Stress

In thin, axisymmetric shells, the stress state is approximated by the membrane state (Fig. 11), where through-thickness variations of stress are negligible. The shell is assumed to be in plane stress, with normal stresses perpendicular to the surface ($\sigma_{nn}$) set to zero.

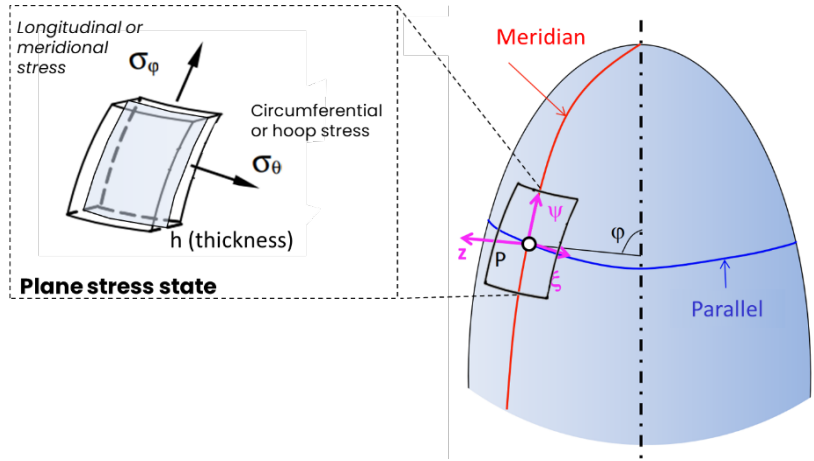

**Fig. 11:** State of stress.

The relevant stress components are:

- $\sigma_{\varphi\varphi}$ – meridional or longitudinal stress (acting along the generating line),
- $\sigma_{\theta\theta}$ – circumferential or hoop stress (acting along the parallels).

The corresponding membrane forces per unit length are defined as reported in Eq. (8). These quantities represent the resultant tangential forces per unit length of the shell in the meridional and circumferential directions, respectively.

$$N_\varphi = \sigma_{\varphi\varphi} h; \quad N_\vartheta = \sigma_{\vartheta\vartheta} h \tag{8}$$

The equilibrium of an infinitesimal shell element under internal pressure $p$ leads to two independent equilibrium equations (neglecting self-weight):

1. equilibrium along the axis of revolution (meridional direction) (Fig. 12 left);
2. equilibrium in the normal direction to the surface (Fig. 12 right).

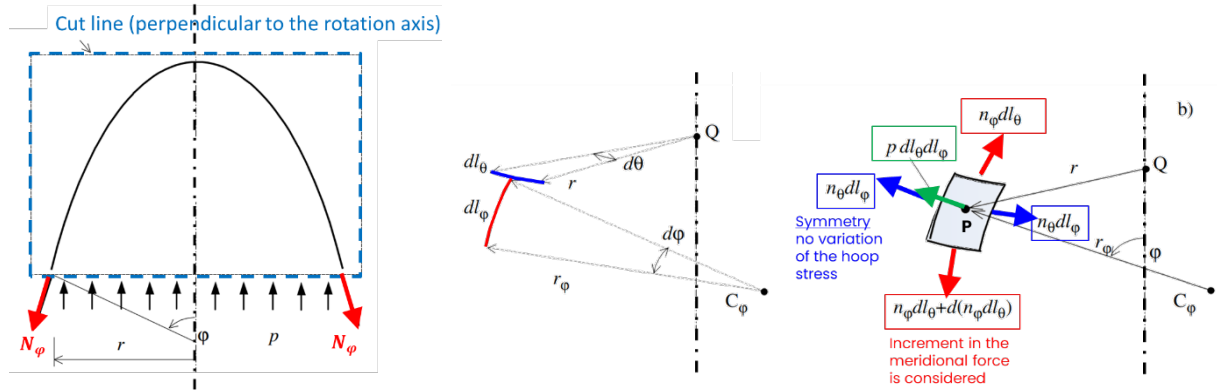

**Fig. 12:** Scheme for equilibrium equations.

These equations, together with boundary conditions, are sufficient to determine $n\varphi$ and $n\theta$. To obtain the first equilibrium equation, the axial equilibrium of a finite portion of the shell must be imposed. This portion is isolated by means of a cut line that interrupts the shell at the section where the meridional force $N_\varphi$ is to be evaluated (Eq. (9)), with a single section perpendicular to the axis. At the section where the vessel is cut, the following forces must be considered. Due to the interruption of the shell wall, the meridional membrane force $N_\varphi$, taken as positive when directed outward. Due to the interruption of the contained fluid, the pressure $p$ (corresponding to the level of the section if the fluid is heavy), taken as positive when directed inward. The equilibrium must also take into account the weight of the fluid mass enclosed within the cut line (but not that located above or below it, outside the isolated portion), whenever it is not negligible—as in the case of gases, it can usually be ignored. The self-weight of the shell wall is generally small compared to the other forces involved and is therefore neglected. In this case, both the self-weight of the vessel and the weight of the contained fluid are disregarded. This operation has to be done once for each change in geometry or loading along the vessel.

$$N_\varphi = \frac{pr}{2\sin\varphi} \tag{9}$$

The second equilibrium equation can be obtained by imposing the translational equilibrium of an infinitesimal shell element in the direction of the normal to the surface. In the analysis of a vessel, once the meridional force $N_\varphi$ has been determined for each portion by applying the procedure described above, the circumferential force $N_\vartheta$ is then obtained by applying Eq. (10).

$$\frac{N_\varphi}{r_\varphi} + \frac{N_\vartheta}{r_\vartheta} = p \tag{10}$$

### 5.1 Examples

#### 5.1.1 Cylindrical Shell under Internal Pressure

Consider a long, thin-walled cylinder of radius $R$, thickness $h$, and internal pressure $p$ (Fig. 13). The equilibrium yields to expression of Eq. (11), from which, dividing by $h$ the corresponding stresses can be obtained. The hoop stress is twice the meridional stress, meaning that yielding typically initiates in the circumferential direction. By applying the Tresca criterion it is possible to determine the minimum thickness for the mechanical resistance of the shell (Eq. (12)), where the admissible tension is the yield stress of the material divided by the safe coefficient.

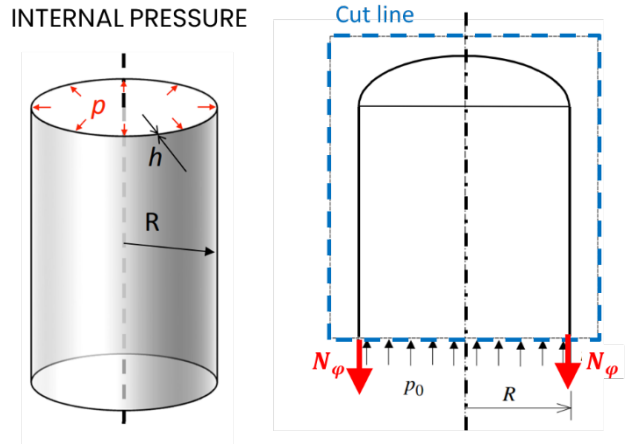

Fig. 13: Cylindrical Shell subjected to Internal Pressure.

$$N_\varphi = \frac{pR}{2}; \; N_\vartheta = pR \; \left(r_\vartheta = R; r_\varphi \to \infty; \varphi = \frac{\pi}{2}\right) \tag{11}$$

$$h_{min} = \frac{N_\vartheta}{\sigma_{adm}} \tag{12}$$

### 5.1.2  *Spherical Shell under Internal Pressure*

Consider a sphere of radius $R$, thickness $h$, and internal pressure $p$ (Fig. 14). The equilibrium yields to expression of Eq. (13), from which, dividing by $h$ the corresponding stresses can be obtained. The stresses are equal in all directions and for each point of the sphere. By applying the Tresca criterion it is possible to determine the minimum thickness for the mechanical resistance of the shell (Eq. (12)).

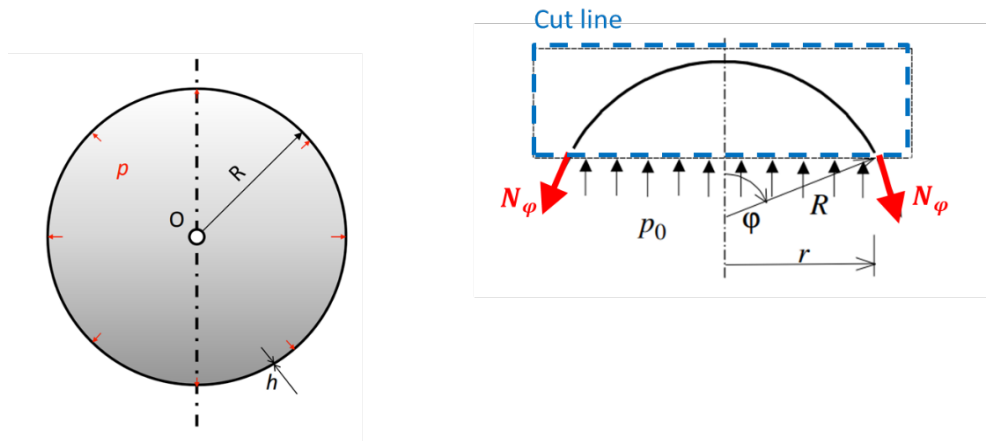

Fig. 14: Spherical shell subjected to internal pressure.

$$N_\varphi = N_\vartheta = \frac{pR}{2} \; (r_\vartheta = r_\varphi = R) \tag{13}$$

### 5.1.3  *Torispherical Heads under Internal Pressure*

By comparing the first two cases presented earlier, it can be observed that, for the same radius and uniform internal pressure, the sphere is subjected to membrane forces whose value is half of the maximum one ($N_\vartheta$) in the cylinder. If the same material and thickness are to be used for all parts, employing a hemispherical cap as the end closure of a cylindrical shell would be rather inefficient, since the material of the sphere would be utilized only up to half of the admissible stress level. Moreover, this type of head would result in greater overall height.

A common design choice (Fig. 15) is therefore to construct the head using a spherical crown of radius $R_s$ equal to twice the cylinder radius $R_c$, which allows the maximum membrane forces to be equalised, connecting it to the cylindrical shell through a toroidal knuckle of small radius $R_t$. A common design choice is to assume $R_t = R_s/10$, and therefore $\sin\varphi_0 \cong 26.4°$. In the toroidal knuckle $r_\varphi = R_t$ and $r_\vartheta = \dfrac{r}{\sin\varphi}$.

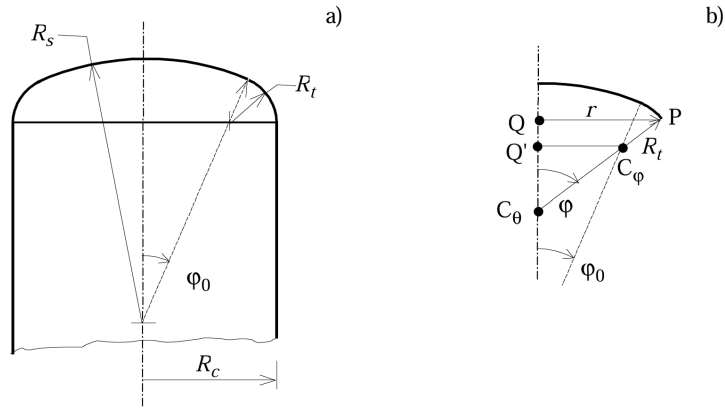

**Fig. 15:** Toro-spherical end and geometrical quantities in the toroidal knuckle.

The torispherical profile exhibits continuity of the tangent and of the curvature radius $r_\vartheta$ at the sphere-to-torus and torus-to-cylinder transition points, but discontinuity of the curvature radius $r_\varphi$, which varies at the first point from Rs to Rt, and at the second point from Rt to infinity. As a consequence, the meridional membrane force $N_\varphi$ is continuous, while the circumferential membrane force $N_\vartheta$ is discontinuous. As in the case of an elliptical head (Fig. 16) where the ratio between the major and minor semi-axes exceed $\sqrt{2}$, in a torispherical head, circumferential compressive hoop stresses are also generated. Such compressive stresses may lead to local buckling or instability phenomena, compromising structural integrity.

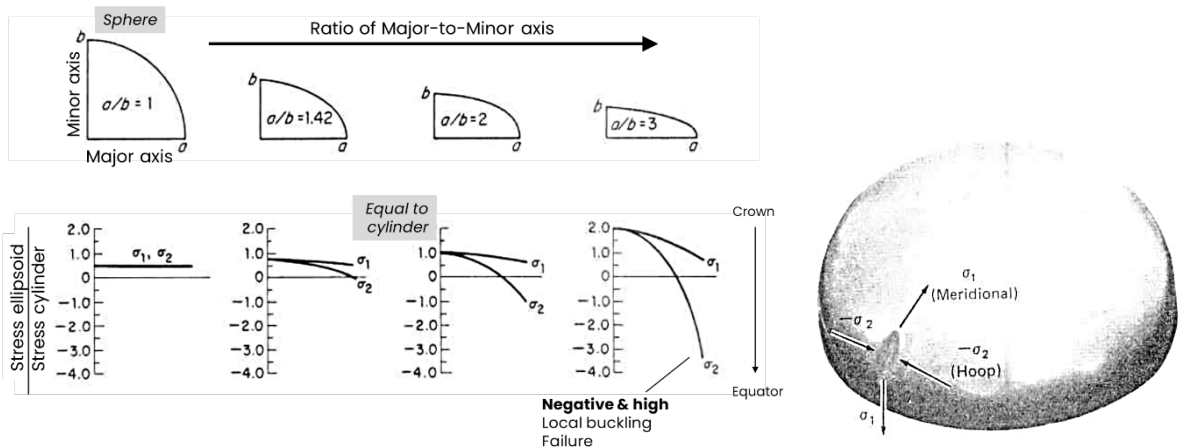

**Fig. 16:** Stress in end with different aspect ratio compared to the cylindrical one (left); buckling in toro-spherical end (right).

## 5.2 Buckling

When a shell is subjected to external pressure (i.e. pressure acting from outside toward the interior), the membrane stresses become compressive. Thin shells are highly sensitive to compressive loading, and instability may occur at stress levels significantly lower than the material's yield stress. The resulting failure is a buckling collapse, often sudden and catastrophic, with little warning. This phenomenon is analogous to the Euler buckling of slender columns under axial compression.

For a cylindrical shell of radius $R$, thickness $h$, and length $L$ under uniform external pressure, elastic instability typically manifests as an ovalization of the cross-section, often forming two lobes (Fig. 17 left). The critical buckling pressure (for two-lobed deformation) is given by Eq. (14).

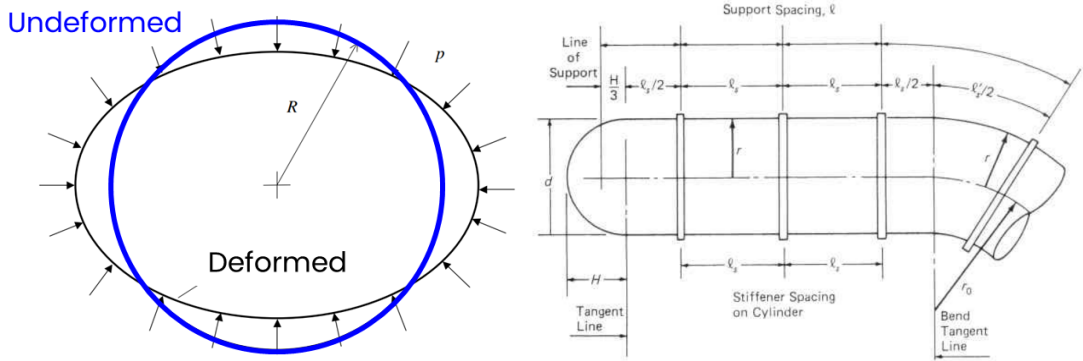

**Fig. 17:** Stress in end with different aspect ratio compared to the cylindrical one (left); buckling in torospherical end (right).

$$p_{cr} = \frac{E}{4(1-\nu^2)}\left(\frac{s}{R}\right)^3 \tag{14}$$

It should also be noted that in the case—always present in practice to a greater or lesser extent—of an initial profile that is not perfectly circular, the critical pressure is still given by Eq. (33); however, the material reaches the elastic limit at pressure values lower than the critical one. Consequently, the critical case is used as a reference, but with high safety factors adopted. If necessary, the instability phenomenon can be counteracted by reinforcing the shell with stiffening rings (Fig. 17 right).

## 6 Secondary (Discontinuity) Stresses

In practice, pressure vessels are composed of different shell portions joined together. At these junctions, geometric or loading discontinuities produce secondary stresses not predicted by pure membrane theory. For instance, at the intersection between a cylindrical body and a hemispherical head, the radial expansion $\Delta R$ under pressure differs for the two components (Fig. 18), in accordance with Eq. (15), which can be applied to each portion. Because the two adjacent parts must remain geometrically continuous, additional bending stresses arise to enforce compatibility of deformation. These edge effects are localized near the junction and decay rapidly away from it. Although they may locally exceed the elastic limit, they do not generally compromise the global integrity of the vessel. For this reason, membrane stresses are called primary, while discontinuity stresses are secondary.

$$\Delta R_i = \varepsilon_{\vartheta_i} R = \frac{1}{E}\left(\sigma_{\vartheta\vartheta_i} - \nu\sigma_{\varphi\varphi_i}\right)R \tag{15}$$

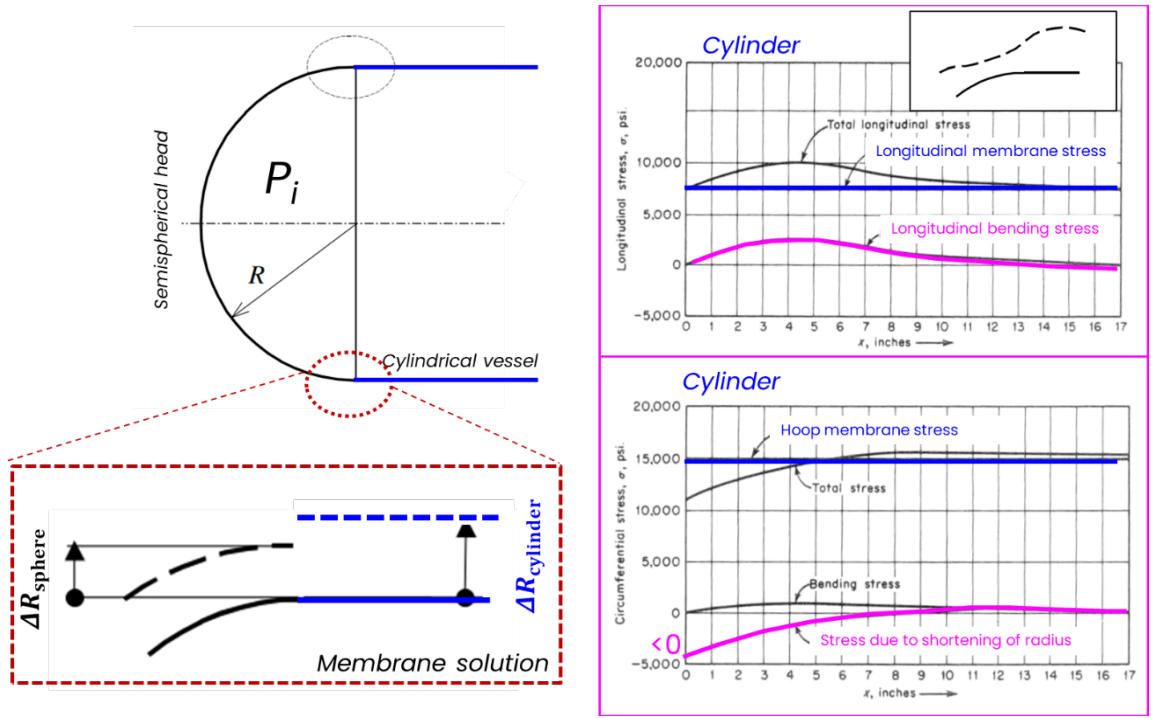

**Fig. 18:** Membrane solution and Secondary stresses.

# 7   Design Considerations and the EN 13445 Standard

The European standard EN 13445 – Unfired Pressure Vessels provides comprehensive design rules for pressure vessels, combining theoretical models and safety requirements. Part 3 of the standard covers design against various failure modes, including yielding, plastic collapse, and buckling. This part provides both the basic rules for defining design conditions and allowable stresses, as well as the calculation methods for a wide range of structural components of pressure vessels.

It is obviously impossible to summarize such an extensive document concisely; in the following, the basic rules and calculation methods adopted for some fundamental components are briefly examined. For further details, reference should be made to the original standard.

In this first section, the basic principles of calculation are defined. The minimum wall thickness values obtained from the calculations must also be ensured by means of a corrosion allowance (when necessary) and by taking into account the tolerances of the semi-finished products. The adopted approach is the "design by formulae" method, meaning that for each component, formulas are provided to determine the required thickness.

The cases to be considered are also defined: operating, exceptional, and test conditions. The presence of welds is taken into account by means of a coefficient, which depends on the non-destructive tests performed. The formula for calculating the admissible stresses are provided depending on the condition and the material.

For a thin shell under internal pressure, the minimum required thickness is derived from the membrane results reported in Chapter 1, but with some modification needed for the presence of welding and to avoid simplifying assumption of plane state of stress. For torispherical head also instability and plasticity are taken into account.

In case of presence of nozzle, a circular opening in the wall of a vessel causes a disturbance in the stress distribution, which must be taken into account in the strength calculation. The effects of the opening are the reduction of the load-bearing cross-section and the stress concentration (notch effect). In the standard, the notch effect is neglected under static loading, relying on the ductility of the material. Therefore, only the reduction of the load-bearing cross-section is considered, corresponding to a rectangle with an area equal to the product of the wall thickness and the diameter of the hole. The calculation procedure requires first verifying whether the opening can be considered isolated from any other nearby openings. The general verification formula for the opening expresses the equilibrium between the pressure acting on the exposed areas and the reaction produced by the stresses in the load-bearing sections.

The calculation of the bolt load and the sizing of the bolts are based on the analysis of the behavior of a flanged joint with a gasket under assembly and operating (or test) conditions. The tube sheets are treated as axisymmetric plates subjected to bending and shear. The shear stress is taken into account (unlike in the case of flat ends) because the axial force tending to tear away the perforated area that accommodates the tube bundle from the outer ring is significant, as it acts on a wall weakened by the tube pass-through holes.

## References

[1] Rösler, J., Harders, H., & Baeker, M. (2007). Mechanical Behaviour of Engineering Materials: Metals, Ceramics, Polymers, and Composites. Springer.

[2] Callister, W. D., & Rethwisch, D. G. (2020). Materials Science and Engineering: An Introduction. Wiley.

[3] Timoshenko, S. P., & Goodier, J. N. (1970). Theory of Elasticity. McGraw-Hill.

[4] Hill, R. (1950). The Mathematical Theory of Plasticity. Oxford University Press.

[5] Chen, W. F., & Han, D. J. (1988). Plasticity for Structural Engineers. Springer.

[6] Harvey, John F. Theory and Design of Pressure Vessels. 3rd ed., Van Nostrand Reinhold, 1985.

[7] Timoshenko, S., and Woinowsky-Krieger, S. (1959). *Theory of Plates and Shells.* McGraw-Hill.

[8] Ugural, A. C. (1981). *Stresses in Plates and Shells.* McGraw-Hill.

[9] EN 13445-3: Unfired Pressure Vessels – Part 3: Design (CEN, latest edition).